\documentclass[aps,prl,twocolumn,showpacs,showdoi,superscriptaddress,superscriptcite]{revtex4}
\usepackage{graphics}
\usepackage{graphicx}
\usepackage{dcolumn}
\usepackage{bm}
\usepackage{float}
\usepackage{subfigure}
\usepackage{todonotes}
\usepackage{color,hyperref}

\begin{document}

\title{Spin reorientation in TlFe$_{1.6}$Se$_2$ with complete vacancy ordering}

\author{Andrew F. May}
\affiliation{Materials Science and Technology Division, Oak Ridge National Laboratory, Oak Ridge, TN 37831}
\author{Michael A. McGuire}
\affiliation{Materials Science and Technology Division, Oak Ridge National Laboratory, Oak Ridge, TN 37831}
\author{Huibo Cao}
\affiliation{Quantum Condensed Matter Division, Oak Ridge National Laboratory, Oak Ridge, TN 37831}
\author{Ilya Sergueev}
\affiliation{European Synchrotron Radiation Facility, Grenoble, France}
\author{Claudia Cantoni}
\affiliation{Materials Science and Technology Division, Oak Ridge National Laboratory, Oak Ridge, TN 37831}
\author{Bryan C. Chakoumakos}
\affiliation{Quantum Condensed Matter Division, Oak Ridge National Laboratory, Oak Ridge, TN 37831}
\author{David S. Parker}
\affiliation{Materials Science and Technology Division, Oak Ridge National Laboratory, Oak Ridge, TN 37831}
\author{Brian C. Sales}
\affiliation{Materials Science and Technology Division, Oak Ridge National Laboratory, Oak Ridge, TN 37831}

\date{\today}

\begin{abstract}
The relationship between vacancy ordering and magnetism in TlFe$_{1.6}$Se$_2$ has been investigated via single crystal neutron diffraction, nuclear forward scattering, and transmission electron microscopy.   The examination of chemically and structurally homogenous crystals allows the true ground state to be revealed, which is characterized by Fe moments lying in the \textit{ab}-plane below 100\,K.  This is in sharp contrast to crystals containing regions of order and disorder, where a competition between \textit{c}-axis and \textit{ab}-plane orientations of the moments is observed.  The properties of partially-disordered TlFe$_{1.6}$Se$_2$ are therefore not associated with solely the ordered or disordered regions.  This contrasts the viewpoint that phase separation results in independent physical properties in intercalated iron selenides, suggesting a coupling between ordered and disordered regions may play an important role in the superconducting analogues.

\end{abstract}

\pacs{74.70.Xa,75.25.-j}
DOI: 10.1103/PhysRevLett.109.077003

\maketitle

The microscopic coexistence of superconductivity and magnetism at temperatures up to $\sim$30\,K in $A_{1-y}$Fe$_{2-x}$Se$_2$ ($A$ = K, Cs, Rb) has generated much interest, and a plethora of techniques have been employed to understand the interesting properties of these materials.\cite{Shermadini2011,Bao2011a,Wen2011Review}  Structural studies on superconducting samples revealed antiferromagnetic order with a local magnetic moment of $\sim$3\,$\mu_B$/Fe, and a $\sqrt{5}a\times\sqrt{5}a$ superstructure of the ThCr$_2$Si$_2$ structure-type derived from Fe vacancy ordering.\cite{Bao2011a,Ye2011}  Other studies have focused on the electronic structure\cite{Cao2011_PRL,Yan2011,Yu2011} and the superconducting pairing mechanism,\cite{Zhang2011_ARPES,Zhang2011,Park2011,Huang2012,Yu2011_NMR,Fang2011_sWave} as well as the influence of phase separation.\cite{Z.Wang2011,Song2011,Ricci2011,Felser2011,Yuan2012,Li2011}

The coexistence of superconductivity and magnetism is now believed to be due to fine-scale phase separation, with each property being associated with a different composition and/or degree of order.\cite{Yuan2012,Ricci2011,Felser2011,Li2011,Ma2011,Charnukha2012}  Phase separation over length scales of $\sim$10-100\,nm has been observed via transmission electron microscopy (TEM),\cite{Z.Wang2011,Song2011,Yuan2012} nano-focused x-ray diffraction,\cite{Ricci2011} and is inferred from M$\ddot{\mathrm{o}}$ssbauer spectroscopy.\cite{Felser2011} Superconducting samples that are single phase (structurally and chemically homoegenous) have yet to be produced and thus the inherent nature of the individual phases remains unclear.  Flexibility in the chemical composition near the parent $A_{0.8}$Fe$_{1.6}$Se$_2$ allows superconductivity to be realized,\cite{Guo2010,Wang2011,Krzton2011,Fang2011,Bao2011b,D.Wang2011} and may do so by promoting phase separation via local inhomogeneity.  It has been suggested that the superconducting phase does not have Fe vacancies.\cite{Li2011,Texier2012,Li2012}  Similar to other Fe-based superconductors, these superconducting phases may be characterized by a doping level of approximately 0.15 electrons per Fe,\cite{Texier2012,PhysRevB.85.140511} such as Rb$_{0.3}$Fe$_2$Se$_2$.  The degree of order is also important,\cite{Han2011,Li2011} however, and first principles calculations have shown that the vacancy and magnetic order greatly influence the Fermi surface.\cite{Yan2011}  

Fe vacancies order into the $\sqrt{5}a\times\sqrt{5}a$ supercell between 460 and 580\,K.  At or slightly below this temperature, the spins align along the tetragonal $c$ axis in a block-checkerboard antiferromagnetic structure (BCAF-$c$, where $c$ indicates the alignment of the spins).\cite{Ye2011,Bao2011a} The vacancy order is generally incomplete, however, with partial occupancies observed on at least one Fe site.\cite{Zavalij2011,Pomjakushin2011}  Superconductivity, which is believed to exist in a phase that does not contain vacancy order, is only observed in samples that also possess regions of vacancy and magnetic order.

In contrast to the alkali-metal based compounds, TlFe$_{1.6}$Se$_2$ has a limited compositional window and the Tl site is always fully occupied, which reduces the degree of inherent disorder.  While superconductivity is observed for mixed Tl / alkali-metal compounds, and was reported with a small volume fraction in TlFe$_{1.7}$Se$_2$,\cite{Fang2011} it has not been reproduced in TlFe$_{1.6}$Se$_2$.  This is perhaps associated with an improper electron count in TlFe$_{1.6}$Se$_2$.  Based on the chemical formula, one would expect TlFe$_{1.6}$Se$_2$ to be a metal.  However, TlFe$_{1.6}$Se$_2$ is observed to be an insulator.  TlFe$_{1.6}$Se$_2$ also displays different magnetic behavior than the alkali-metal analogues.

The BCAF-$c$ structure is the only magnetic structure observed in the alkali-metal compounds.  However, single crystals of `partially-disordered TlFe$_{1.6}$Se$_2$' have been shown to possess additional magnetic phase transitions near 100\,K and 140\,K.\cite{SalesTFS}  A detailed structural study revealed these transitions are associated with a canting of the Fe moments toward the $ab$-plane, which only occurs between $\sim$100 and 150\,K.\cite{CaoTFS}  These crystals are characterized by crystallographically coherent regions of ordered and disordered Fe vacancies.\cite{CaoTFS}  In the absence of single-phase crystals, it is impossible to know if these magnetic transitions in partially-disordered TlFe$_{1.6}$Se$_2$ are inherent to the ordered or disordered regions, or are the result of some interaction between the two.

In this Letter, we provide a detailed characterization of single crystal TlFe$_{1.6}$Se$_2$ with complete chemical/vacancy order.  Such structural and chemical homogeneity has been elusive in $A_{1-y}$Fe$_{2-x}$Se$_2$, and the presence of multiple phases has complicated the structure-property investigations in these complex materials.  By obtaining ideal order, a previously unobserved spin reorientation is revealed with spins lying in the $\textit{ab}$-plane for $T <$ 100\,K.  This suggests a strong interaction between the ordered and disordered regions prevents this ground state from occurring in the partially-disordered crystals at low temperatures.

Crystals with complete vacancy order were grown in melts of nominal composition TlFe$_{1.6}$Se$_2$.  In the same furnace, a nominal composition of TlFe$_{1.7}$Se$_2$ produced crystals with magnetization behavior qualitatively similar to the `partially disordered' crystals, which were grown from a melt of nominal composition TlFe$_{2}$Se$_2$.\cite{SalesTFS}  Therefore, growth in iron-rich environments appears to inhibit the ordering of iron vacancies, perhaps due to minor differences in Fe content that are difficult to detect.  In the fully ordered crystals, the vacancy ordering temperature is found to be $\sim$463\,K by differential scanning calorimetry (DSC), and an anomaly in the magnetic susceptibility ($\chi$) was also observed at this temperature.  An entropy change of $\sim$7.4\,J/mol-Fe/K was calculated from the DSC data, and this is $\sim$75\% of the entropy change expected for the simultaneous ordering of Fe vacancies and Fe magnetic moments.  See the \href{http://prl.aps.org/supplemental/PRL/v109/i7/e077003}{Supplemental Information} for additional details.

\begin{figure}
	\centering
\includegraphics[width=2.8in]{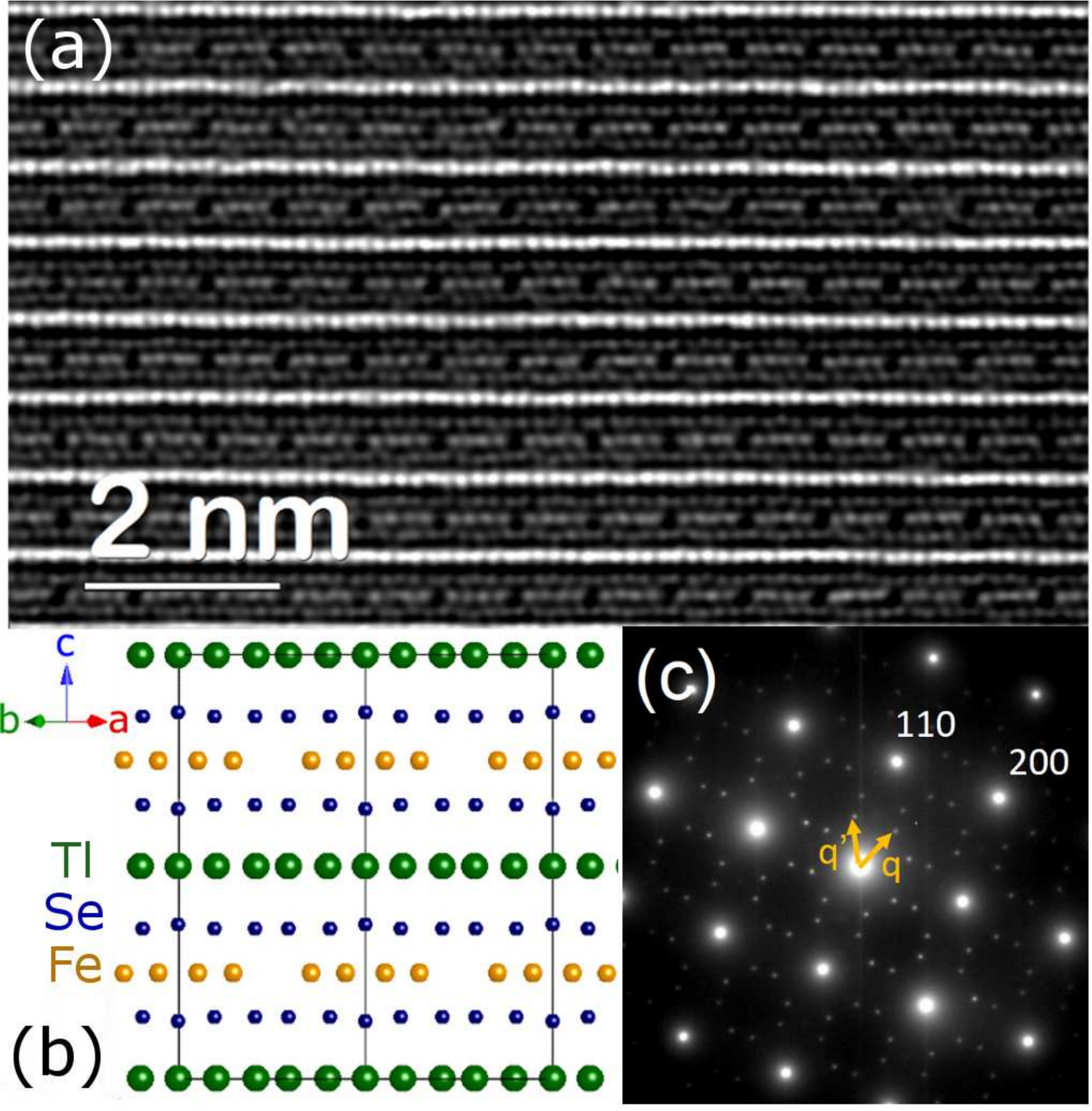}
\caption{(color online) (a) Aberration corrected HAADF image of a chemically homogenous crystal, with electron beam along the [110] direction of the $\sqrt{5}a\times\sqrt{5}a$ supercell, where all Fe vacancies are aligned in columns viewed as the dark spots.  A corresponding crystal structure drawing is shown in (b). (c) Selected area diffraction with electron beam along [001] showing reflections from the $\sqrt{5}a\times\sqrt{5}a$ superstructure with twinning yielding spots from two domains indicated by q and q'; the pattern is indexed according to the subcell.}
	\label{fig:tem}
\end{figure}

The crystals were observed to be chemically homogenous and fully-ordered via aberration corrected scanning TEM (STEM) using high angular annular dark field (HAADF) imaging.  The STEM image shown in Fig.\,\ref{fig:tem}a was taken with the electron beam parallel to the [110] direction of the supercell.  As a result of this orientation, all of the Fe vacancies are aligned in columns parallel to the electron beam and can be readily viewed as dark spots separated by four iron columns.  A crystal structure drawing corresponding to the HAADF image is shown in Fig.\,\ref{fig:tem}b to highlight this ideal vacancy ordering, and the atomic layers are labeled for ease of viewing.  Specimens viewed along [001] also revealed an ordered and uniform vacancy superstructure, and a selected area diffraction pattern from this orientation is shown in Fig.\,\ref{fig:tem}c.  The selected area diffraction pattern shows two sets of $\sqrt{5}a\times\sqrt{5}a$ superstructure reflections, indicated by the vectors q and q', due to twinning along the (110) subcell planes.  Images of the partially-disordered crystals can be found in Ref.\,\citenum{CaoTFS}. STEM data reveal the disordered regions contain Fe vacancies (modulations in Fe HAADF intensity), and thus these regions are not like the vacancy-free phases being reported in the superconducting analogues.

\begin{figure}
	\centering
\includegraphics[width=3in]{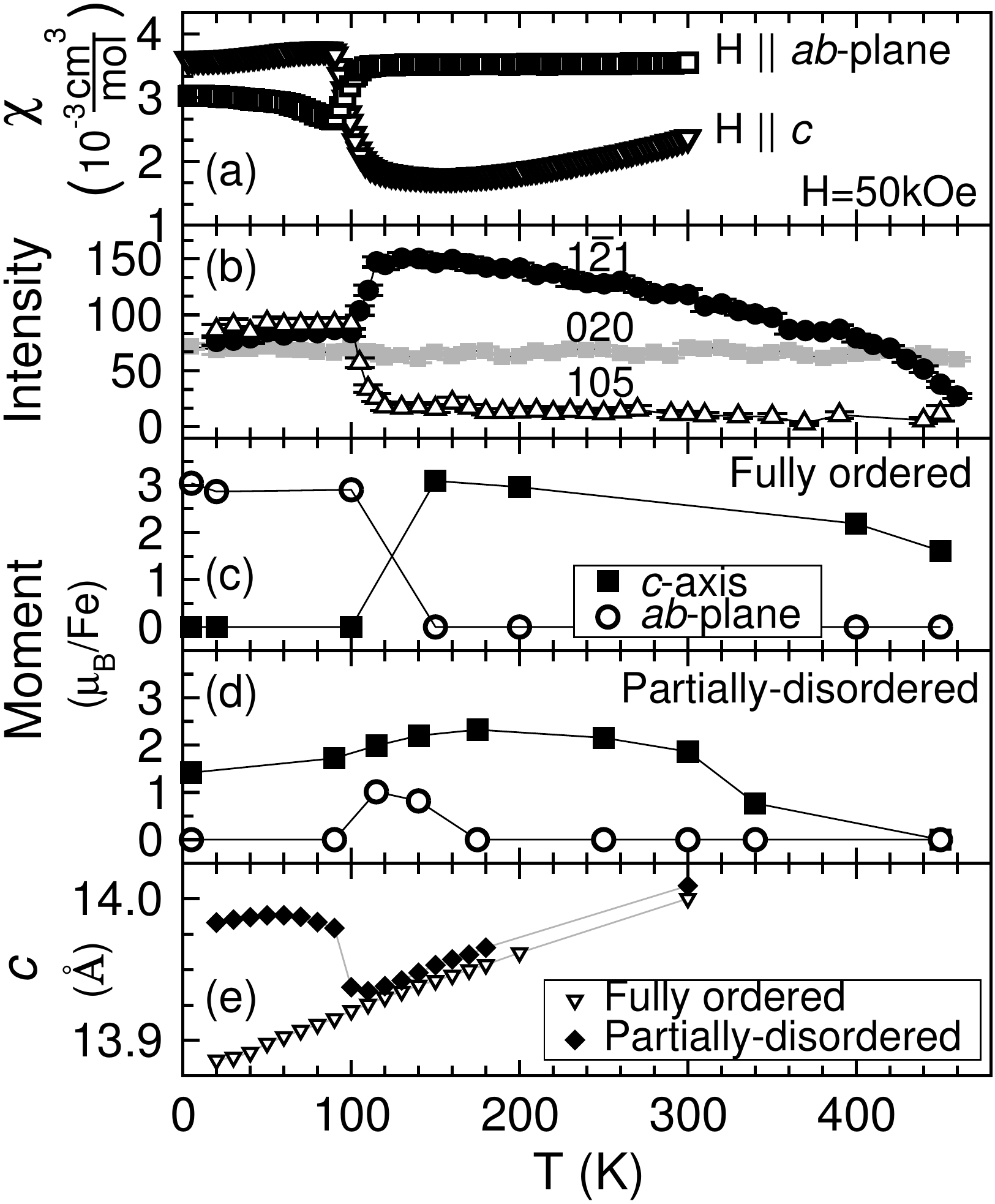}
\caption{(a) Magnetic susceptibility of oriented single crystals reveals a change in the easy magnetic axis near 100\,K. (b) The integrated neutron diffraction intensity (in counts) for the vacancy-order (020) peak demonstrates a saturation of the vacancy ordering at high temperature.  The integrated intensity of the (1$\bar{2}$1) is proportional to the $c$-axis component of the moment, and the (105) peak gains intensity when the moment has an $ab$-plane component; together, these magnetic peaks reveal a change in magnetic orientation near 100\,K.  (c) Projections of the refined moments onto the $ab$-plane and $c$-axis for fully ordered crystals and in (d) the same is shown for partially-ordered crystals. (e) The $c$-lattice parameters obtained by x-ray diffraction.}
	\label{fig:mag}
\end{figure}

Refinements of the neutron diffraction data indicate complete occupancy of the Fe1 site (Wyckoff position 16$i$) and zero occupation of the Fe2 site (Wyckoff position 4$d$), which yields the ideal $\sqrt{5}a\times\sqrt{5}a$ superstructure in TlFe$_{1.6}$Se$_2$.  In Fig.\,\ref{fig:mag}b, the intensity of the vacancy order (020) peak is shown to be independent of temperature below $\sim$460\,K, indicating a saturation of vacancy order at all temperatures probed by neutron diffraction. 

The magnetic susceptibility of fully ordered TlFe$_{1.6}$Se$_2$ single crystals is shown in Fig.\,\ref{fig:mag}a, where a phase transition is clearly observed near 100\,K.  Above this transition, the susceptibility is largest when the magnetic field (\textbf{H}) is perpendicular to \textit{c}, and for temperatures below the transition the susceptibility is largest for \textbf{H} $\parallel$ \textit{c}.  This indicates a switch in the easy magnetic direction, with moments parallel to \textit{c} above  $\sim$100\,K and perpendicular to \textit{c} below $\sim$100\,K.  

\begin{figure}
\centering
\includegraphics[width=2.85in]{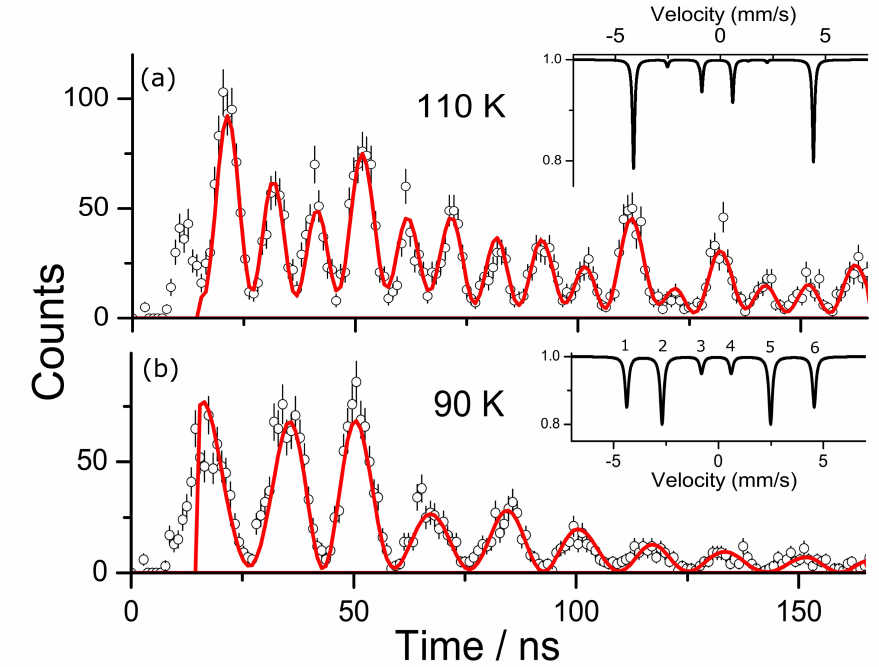}
\caption{(color online) (color online) Time evolution of NFS for fully ordered TlFe$_{1.6}$Se$_2$ crystal with the $c$-axis parallel to the x-ray beam measured above and below the magnetic transition. The data are fit by models (red lines) with magnetic moments aligned with the $c$-axis at 110\,K, and lying in the $ab$-plane at 90\,K. The insets are the model simulations of M$\ddot{\mathrm{o}}$ssbauer spectra, with the lines numbered in the inset of (b).}
\label{fig:nfs}
\end{figure}

The reorientation of the magnetic moment is confirmed via single crystal neutron diffraction (Fig.\,\ref{fig:mag}b,c) and synchrotron radiation based nuclear forward scattering (Fig.\,\ref{fig:nfs}).  Figure \ref{fig:mag}b shows the integrated intensity for three Bragg peaks observed via neutron diffraction on a $\sim$50\,mg single crystal.  The intensities of the two peaks with magnetic contributions change rapidly near 100\,K.  The (1$\bar{2}$1) peak intensity is associated with moments aligned with the $c$-axis and is beginning to saturate before it quickly diminishes near 100\,K.  At this temperature, the intensity of the (105) peak increases sharply due to the onset of the new magnetic order.  The (105) peak gains intensity when the $ab$-plane component of the moment increases, as well as when the Fe layer develops a corrugation.\cite{CaoTFS}  The moment reorientation is highlighted in Fig.\,\ref{fig:mag}c where the refined moments are presented, and reach $\sim$3$\mu_B$/Fe in both magnetic structures.  A similar moment is obtained via first principles calculations for the BCAF-$c$ (\href{http://prl.aps.org/supplemental/PRL/v109/i7/e077003}{Supplemental Information}), and the current calculations are consistent with previous reports.\cite{Yan2011}

The canting of the Fe moment in the partially-disordered TlFe$_{1.6}$Se$_2$ between 100 and 150\,K is shown in Figure \ref{fig:mag}d.\cite{CaoTFS} A competition between the BCAF-$c$ and an in-plane structure exists in the partially-disordered crystals, where the BCAF-$c$ structure is present below 100\,K and above 150\,K, and thus the corresponding moment has no $ab$-plane component.  When this canting is lost below $\sim$100\,K, an abrupt increase in the $c$-lattice parameter is observed (see Fig. \ref{fig:mag}e).\cite{SalesTFS,CaoTFS}  The magnetic transition in the fully ordered crystals is not associated with an anomaly in the $c$ lattice parameter, though, and this is consistent with the smooth behavior in $c$ near $\sim$150\,K where the canting of the moment begins in the partially-disordered crystals.

The moment reorientation is also observed via nuclear forward scattering (NFS),\cite{Rohlsberger} the time-analogue of Mossbauer spectroscopy.  NFS measurements were completed at beam line ID18 of the European Synchrotron Radiation Facility,\cite{Ruffer} with the crystallographic $c$-axis parallel to the x-ray beam.  The spectra were fit by a conventional routine.\cite{MOTIF}

The NFS data at temperatures slightly above and below magnetic transition at 100\,K are shown in Fig.\,\ref{fig:nfs}. These results clearly indicate the change of the hyperfine structure across the transition, and confirm the moment orientation changes from parallel to $c$ above 100\,K to perpendicular to $c$ below 100\,K. At 110\,K, the data are well described by a model with AFM Fe moments aligned with the $c$-axis, yielding a magnetic field H=26.1\,T, a quadrupole splitting  of $\Delta$E$_{\textrm{Q}}$=1.18\,mm/s, and the angle between main axes of those interactions is 46$^{\circ}$.  These values are in excellent agreement with those observed for Rb$_{0.8}$Fe$_{1.6}$Se$_2$.\cite{Felser2011}

The NFS data at 90\,K were fitted with 2 Fe sites with moments aligned in $ab$-plane, and the angle between them is utilized as a fit parameter.  This model can generally describe any complex, non-collinear magnetic structure with moments in the plane.  Independent of the model utilized, a change of the carrier frequency from that related to the beats of 1$^{\textrm{st}}$ and 6$^{\textrm{th}}$ lines of Mossbauer spectrum (at 110\,K) to that related to the beats of 2$^{\textrm{nd}}$ and 5$^{\textrm{th}}$ lines (at 90\,K) is observed, indicating a reorientation of the Fe moments.  An appearance of the strong 2$^{\textrm{nd}}$ and 5$^{\textrm{th}}$ lines indicates the $ab$-plane alignment of the Fe magnetic moments below 100\,K.  See the \href{http://prl.aps.org/supplemental/PRL/v109/i7/e077003}{Supplemental Information} for further details on NFS data collection and analysis.

\begin{figure}
\centering
\includegraphics[width=3.2in]{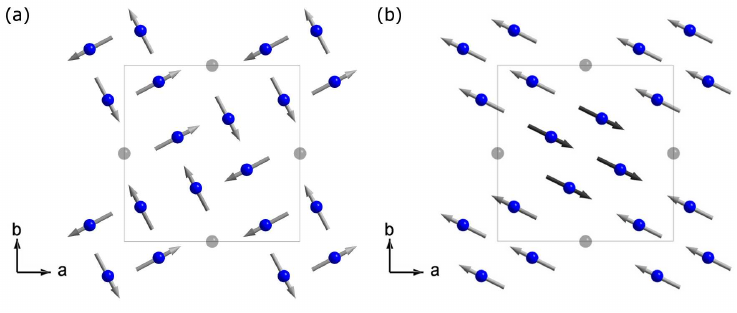}
\caption{(color online) The Fe (blue) lattice with differernt magnetic structures that describe the neutron diffraction data below 100\,K. (a) A non-collinear magnetic structure that describes the neutron diffraction data and retains the symmetry of the crystal structure.  (b) An in-plane, block-checker antiferromagnetic structure that describes the neutron diffraction data provided an equal mixture of domains rotated 90$^{\circ}$ about $c$ exists. In both (a) and (b), the moments (arrows) are antiferromagnetically aligned between Fe planes along the $c$-axis, and vacant Fe sites are shown as light grey.}
\label{fig:spins}
\end{figure}

Below 100\,K, the neutron diffraction data were modeled using several different magnetic structures.  First, the eight possible magnetic structures within the expected tetragonal symmetry were examined.  Of these, the best fit ($R$=0.052 at 5\,K) corresponds to the non-collinear magnetic structure ($I$4/m) shown in Fig.\,\ref{fig:spins}a with moments pointing 10.8$^{\circ}$ away from the vacancies. Fixing the moments towards the vacancies results in a similar quality fit.

Alternative magnetic models require lowering the symmetry of the magnetic structure and assuming equal twin fractions of collinear magnetic domains. The monoclinic model ($I$2'/m) shown in Fig.\,\ref{fig:spins}b (and a twin generated by rotating 90 degrees about the $c$-axis) can describe the data well ($R$ = 0.045 with 2.99(4)$\mu_B$/Fe, at 5\,K). Other possible twinned collinear magnetic structure models were tested but no others gave reasonable agreement.

Theoretical calculations have suggested non-collinear structures are stable in the vacancy-ordered systems.\cite{Yu2011} Collinear, in-plane structures are observed in the majority of Fe-based superconducting families,\cite{Wen2011Review,Lumsden_Christianson_Review} but the collinear structure shown in Fig.\,\ref{fig:spins}b would be unique to this system.  The measurements presented here are incapable of conclusively identifying the magnetic structure below 100\,K.  In the Fe-based superconductors, structural and magnetic transitions are typically coupled.  Therefore, it seems reasonable to expect that if the magnetic symmetry were lowered to monoclinic, to allow the structure in Fig.\,\ref{fig:spins}b, then the nuclear structure would also distort.  Such a distortion is not evident in the current diffraction data.  As such, the preference would typically be for the higher symmetry structure (Fig.\,\ref{fig:spins}a), which also provides a good description of the data. Regardless of the exact spin structure, this reorientation of the spins near 100\,K is clearly driven by the complete ordering of vacancies, and the dominance of this ground state is unique to this phase-pure material.

The transitions at 100\,K in the fully-ordered and 140\,K in the partially-disordered crystals are similar.  In both cases, the moments develop an $ab$-plane component upon cooling through the transition.  The structures in Fig.\,\ref{fig:spins} were not observed via neutron diffraction for the partially-disordered crystals, though the NFS data suggest the Fe moments lie within the $ab$-plane at 110\,K (see \href{http://prl.aps.org/supplemental/PRL/v109/i7/e077003}{Supplemental Information}).  In either case, the BCAF-$c$ order with spins along $c$ is the ground state in the partially-disordered crystals below 100\,K.  Therefore, the disordered regions, which are likely to be non-magnetic,\cite{Felser2011} have a strong influence on the observed magnetic behavior.  Consistent with these results, it has been recently suggested that superconductivity is induced in KFe$_2$Se$_2$ when this KFe$_2$Se$_2$ stoichiometry interacts with the vacancy-ordered K$_{0.8}$Fe$_{1.6}$Se$_2$ phase.\cite{Li2012}

Given the magnetoelastic nature of these materials,\cite{ZhangSingh2009,SalesTFS,CaoTFS} it is not surprising that disordered and ordered (non-magnetic and magnetic) regions would be strongly coupled. These regions are crystallographically coherent and approximately 10-20\,nm in $ab$-plane and 2-4 unit cells along $c$,\cite{CaoTFS} which allows them to interact via lattice strain.   A detailed analysis of the STEM data for partially-disordered TlFe$_{1.6}$Se$_2$ reveals the $c$ lattice parameter is approximately 2\% larger in the disordered regions than in the ordered regions, and the change in $a$ is less than $\sim$0.3\%.  Therefore, modulations in $c$ are the most likely source for the strain interaction. 

As demonstrated in this Letter, the absence of disordered regions allows TlFe$_{1.6}$Se$_2$ to complete a spin reorientation at $\sim$100\,K, thereby unveiling the true magnetic ground state associated with the $\sqrt{5}a\times\sqrt{5}a$ supercell.  This demonstrates a strong influence of the secondary phase on the behavior of the primary phase, and thus an interaction between the ordered and disordered regions determines the properties of the partially-disordered crystals.  These results have implications for the superconducting alkali-metal iron selenides, which also possess regions of order and disorder separated on a fine scale.  Understanding the interaction between these regions may prove critical in identifying the true nature of superconductivity in those materials.

Research was supported in part by the Materials Sciences and Engineering Division, Office of Science, US Department of Energy (A.F.M, C.C., M.A.M., B.C.S.). D.S.P. was supported by the ORNL LDRD SEED funding project S12-006, ''Rare Earth Free Magnets: Compute, Create, Characterize.'' The research at ORNL's High Flux Isotope Reactor was sponsored by the Scientific User Facilities Division, Office of Basic Energy Sciences, US Department of Energy. This research was also partially supported by ORNL SHaRE, sponsored by the Division of Scientific User Facilities, Office of Basic Energy Sciences, U.S. Department of Energy. The European Synchrotron Radiation Facility is acknowledged for provision of synchrotron radiation facilities at beam line ID18.


\begin{thebibliography}{40}
\expandafter\ifx\csname natexlab\endcsname\relax\def\natexlab#1{#1}\fi
\expandafter\ifx\csname bibnamefont\endcsname\relax
  \def\bibnamefont#1{#1}\fi
\expandafter\ifx\csname bibfnamefont\endcsname\relax
  \def\bibfnamefont#1{#1}\fi
\expandafter\ifx\csname citenamefont\endcsname\relax
  \def\citenamefont#1{#1}\fi
\expandafter\ifx\csname url\endcsname\relax
  \def\url#1{\texttt{#1}}\fi
\expandafter\ifx\csname urlprefix\endcsname\relax\def\urlprefix{URL }\fi
\providecommand{\bibinfo}[2]{#2}
\providecommand{\eprint}[2][]{\url{#2}}

\bibitem[{\citenamefont{Shermadini et~al.}(2011)\citenamefont{Shermadini,
  Krzton-Maziopa, Bendele, Khasanov, Luetkens, Conder, Pomjakushina, Weyeneth,
  Pomjakushin, Bossen et~al.}}]{Shermadini2011}
\bibinfo{author}{\bibfnamefont{Z.}~\bibnamefont{Shermadini}},
  \bibinfo{author}{\bibfnamefont{A.}~\bibnamefont{Krzton-Maziopa}},
  \bibinfo{author}{\bibfnamefont{M.}~\bibnamefont{Bendele}},
  \bibinfo{author}{\bibfnamefont{R.}~\bibnamefont{Khasanov}},
  \bibinfo{author}{\bibfnamefont{H.}~\bibnamefont{Luetkens}},
  \bibinfo{author}{\bibfnamefont{K.}~\bibnamefont{Conder}},
  \bibinfo{author}{\bibfnamefont{E.}~\bibnamefont{Pomjakushina}},
  \bibinfo{author}{\bibfnamefont{S.}~\bibnamefont{Weyeneth}},
  \bibinfo{author}{\bibfnamefont{V.}~\bibnamefont{Pomjakushin}},
  \bibinfo{author}{\bibfnamefont{O.}~\bibnamefont{Bossen}},
  \bibnamefont{et~al.}, \bibinfo{journal}{Phys. Rev. Lett.}
  \textbf{\bibinfo{volume}{106}}, \bibinfo{pages}{117602}
  (\bibinfo{year}{2011}),
  \urlprefix\url{http://link.aps.org/doi/10.1103/PhysRevLett.106.117602}.

\bibitem[{\citenamefont{Bao et~al.}(2011)\citenamefont{Bao, Huang, Chen, Green,
  Wang, He, and Qiu}}]{Bao2011a}
\bibinfo{author}{\bibfnamefont{W.}~\bibnamefont{Bao}},
  \bibinfo{author}{\bibfnamefont{Q.}~\bibnamefont{Huang}},
  \bibinfo{author}{\bibfnamefont{G.~F.} \bibnamefont{Chen}},
  \bibinfo{author}{\bibfnamefont{M.~A.} \bibnamefont{Green}},
  \bibinfo{author}{\bibfnamefont{D.~M.} \bibnamefont{Wang}},
  \bibinfo{author}{\bibfnamefont{J.~B.} \bibnamefont{He}}, \bibnamefont{and}
  \bibinfo{author}{\bibfnamefont{Y.}~\bibnamefont{Qiu}},
  \bibinfo{journal}{Chinese Phys. Lett.} \textbf{\bibinfo{volume}{28}},
  \bibinfo{pages}{086104} (\bibinfo{year}{2011}).

\bibitem[{\citenamefont{Wen et~al.}(2011)\citenamefont{Wen, Xu, Gu, Tranquada,
  and Birgeneau}}]{Wen2011Review}
\bibinfo{author}{\bibfnamefont{J.}~\bibnamefont{Wen}},
  \bibinfo{author}{\bibfnamefont{G.}~\bibnamefont{Xu}},
  \bibinfo{author}{\bibfnamefont{G.}~\bibnamefont{Gu}},
  \bibinfo{author}{\bibfnamefont{J.~M.} \bibnamefont{Tranquada}},
  \bibnamefont{and} \bibinfo{author}{\bibfnamefont{R.~J.}
  \bibnamefont{Birgeneau}}, \bibinfo{journal}{Rep. Prog. Phys.}
  \textbf{\bibinfo{volume}{74}}, \bibinfo{pages}{124503}
  (\bibinfo{year}{2011}).

\bibitem[{\citenamefont{Ye et~al.}(2011)\citenamefont{Ye, Chi, Bao, Wang, Ying,
  Chen, Wang, Dong, and Fang}}]{Ye2011}
\bibinfo{author}{\bibfnamefont{F.}~\bibnamefont{Ye}},
  \bibinfo{author}{\bibfnamefont{S.}~\bibnamefont{Chi}},
  \bibinfo{author}{\bibfnamefont{W.}~\bibnamefont{Bao}},
  \bibinfo{author}{\bibfnamefont{X.~F.} \bibnamefont{Wang}},
  \bibinfo{author}{\bibfnamefont{J.~J.} \bibnamefont{Ying}},
  \bibinfo{author}{\bibfnamefont{X.~H.} \bibnamefont{Chen}},
  \bibinfo{author}{\bibfnamefont{H.~D.} \bibnamefont{Wang}},
  \bibinfo{author}{\bibfnamefont{C.~H.} \bibnamefont{Dong}}, \bibnamefont{and}
  \bibinfo{author}{\bibfnamefont{M.}~\bibnamefont{Fang}},
  \bibinfo{journal}{Phys. Rev. Lett.} \textbf{\bibinfo{volume}{107}},
  \bibinfo{pages}{137003} (\bibinfo{year}{2011}),
  \urlprefix\url{http://link.aps.org/doi/10.1103/PhysRevLett.107.137003}.

\bibitem[{\citenamefont{Cao and Dai}(2011)}]{Cao2011_PRL}
\bibinfo{author}{\bibfnamefont{C.}~\bibnamefont{Cao}} \bibnamefont{and}
  \bibinfo{author}{\bibfnamefont{J.}~\bibnamefont{Dai}},
  \bibinfo{journal}{Phys. Rev. Lett.} \textbf{\bibinfo{volume}{107}},
  \bibinfo{pages}{056401} (\bibinfo{year}{2011}),
  \urlprefix\url{http://link.aps.org/doi/10.1103/PhysRevLett.107.056401}.

\bibitem[{\citenamefont{Yan et~al.}(2011)\citenamefont{Yan, Gao, Lu, and
  Xiang}}]{Yan2011}
\bibinfo{author}{\bibfnamefont{X.-W.} \bibnamefont{Yan}},
  \bibinfo{author}{\bibfnamefont{M.}~\bibnamefont{Gao}},
  \bibinfo{author}{\bibfnamefont{Z.-Y.} \bibnamefont{Lu}}, \bibnamefont{and}
  \bibinfo{author}{\bibfnamefont{T.}~\bibnamefont{Xiang}},
  \bibinfo{journal}{Phys. Rev. B} \textbf{\bibinfo{volume}{83}},
  \bibinfo{pages}{233205} (\bibinfo{year}{2011}),
  \urlprefix\url{http://link.aps.org/doi/10.1103/PhysRevB.83.233205}.

\bibitem[{\citenamefont{Yu et~al.}(2011{\natexlab{a}})\citenamefont{Yu,
  Goswami, and Si}}]{Yu2011}
\bibinfo{author}{\bibfnamefont{R.}~\bibnamefont{Yu}},
  \bibinfo{author}{\bibfnamefont{P.}~\bibnamefont{Goswami}}, \bibnamefont{and}
  \bibinfo{author}{\bibfnamefont{Q.}~\bibnamefont{Si}}, \bibinfo{journal}{Phys.
  Rev. B} \textbf{\bibinfo{volume}{84}}, \bibinfo{pages}{094451}
  (\bibinfo{year}{2011}{\natexlab{a}}),
  \urlprefix\url{http://link.aps.org/doi/10.1103/PhysRevB.84.094451}.

\bibitem[{\citenamefont{Zhang et~al.}(2011{\natexlab{a}})\citenamefont{Zhang,
  Yang, Xu, Ye, Chen, He, Xu, Jiang, Xie, Ying et~al.}}]{Zhang2011_ARPES}
\bibinfo{author}{\bibfnamefont{Y.}~\bibnamefont{Zhang}},
  \bibinfo{author}{\bibfnamefont{L.~X.} \bibnamefont{Yang}},
  \bibinfo{author}{\bibfnamefont{M.}~\bibnamefont{Xu}},
  \bibinfo{author}{\bibfnamefont{Z.~R.} \bibnamefont{Ye}},
  \bibinfo{author}{\bibfnamefont{F.}~\bibnamefont{Chen}},
  \bibinfo{author}{\bibfnamefont{C.}~\bibnamefont{He}},
  \bibinfo{author}{\bibfnamefont{H.~C.} \bibnamefont{Xu}},
  \bibinfo{author}{\bibfnamefont{J.}~\bibnamefont{Jiang}},
  \bibinfo{author}{\bibfnamefont{B.~P.} \bibnamefont{Xie}},
  \bibinfo{author}{\bibfnamefont{J.~J.} \bibnamefont{Ying}},
  \bibnamefont{et~al.}, \bibinfo{journal}{Nature Mater.}
  \textbf{\bibinfo{volume}{10}}, \bibinfo{pages}{273}
  (\bibinfo{year}{2011}{\natexlab{a}}).

\bibitem[{\citenamefont{Zhang et~al.}(2011{\natexlab{b}})\citenamefont{Zhang,
  Lu, and Xiang}}]{Zhang2011}
\bibinfo{author}{\bibfnamefont{G.~M.} \bibnamefont{Zhang}},
  \bibinfo{author}{\bibfnamefont{Z.~Y.} \bibnamefont{Lu}}, \bibnamefont{and}
  \bibinfo{author}{\bibfnamefont{T.}~\bibnamefont{Xiang}},
  \bibinfo{journal}{Phys. Rev. B} \textbf{\bibinfo{volume}{84}},
  \bibinfo{pages}{052502} (\bibinfo{year}{2011}{\natexlab{b}}),
  \urlprefix\url{http://link.aps.org/doi/10.1103/PhysRevB.84.052502}.

\bibitem[{\citenamefont{Park et~al.}(2011)\citenamefont{Park, Friemel, Li, Kim,
  Tsurkan, Deisenhofer, Krug~von Nidda, Loidl, Ivanov, Keimer
  et~al.}}]{Park2011}
\bibinfo{author}{\bibfnamefont{J.~T.} \bibnamefont{Park}},
  \bibinfo{author}{\bibfnamefont{G.}~\bibnamefont{Friemel}},
  \bibinfo{author}{\bibfnamefont{Y.}~\bibnamefont{Li}},
  \bibinfo{author}{\bibfnamefont{J.-H.} \bibnamefont{Kim}},
  \bibinfo{author}{\bibfnamefont{V.}~\bibnamefont{Tsurkan}},
  \bibinfo{author}{\bibfnamefont{J.}~\bibnamefont{Deisenhofer}},
  \bibinfo{author}{\bibfnamefont{H.-A.} \bibnamefont{Krug~von Nidda}},
  \bibinfo{author}{\bibfnamefont{A.}~\bibnamefont{Loidl}},
  \bibinfo{author}{\bibfnamefont{A.}~\bibnamefont{Ivanov}},
  \bibinfo{author}{\bibfnamefont{B.}~\bibnamefont{Keimer}},
  \bibnamefont{et~al.}, \bibinfo{journal}{Phys. Rev. Lett.}
  \textbf{\bibinfo{volume}{107}}, \bibinfo{pages}{177005}
  (\bibinfo{year}{2011}),
  \urlprefix\url{http://link.aps.org/doi/10.1103/PhysRevLett.107.177005}.

\bibitem[{\citenamefont{Huang and Mou}(2012)}]{Huang2012}
\bibinfo{author}{\bibfnamefont{S.~M.} \bibnamefont{Huang}} \bibnamefont{and}
  \bibinfo{author}{\bibfnamefont{C.~Y.} \bibnamefont{Mou}},
  \bibinfo{journal}{Phys. Rev. B} \textbf{\bibinfo{volume}{85}},
  \bibinfo{pages}{184519} (\bibinfo{year}{2012}).

\bibitem[{\citenamefont{Yu et~al.}(2011{\natexlab{b}})\citenamefont{Yu, Ma, He,
  Wang, Xia, Chen, and Bao}}]{Yu2011_NMR}
\bibinfo{author}{\bibfnamefont{W.}~\bibnamefont{Yu}},
  \bibinfo{author}{\bibfnamefont{L.}~\bibnamefont{Ma}},
  \bibinfo{author}{\bibfnamefont{J.~B.} \bibnamefont{He}},
  \bibinfo{author}{\bibfnamefont{D.~M.} \bibnamefont{Wang}},
  \bibinfo{author}{\bibfnamefont{T.-L.} \bibnamefont{Xia}},
  \bibinfo{author}{\bibfnamefont{G.~F.} \bibnamefont{Chen}}, \bibnamefont{and}
  \bibinfo{author}{\bibfnamefont{W.}~\bibnamefont{Bao}},
  \bibinfo{journal}{Phys. Rev. Lett.} \textbf{\bibinfo{volume}{106}},
  \bibinfo{pages}{197001} (\bibinfo{year}{2011}{\natexlab{b}}),
  \urlprefix\url{http://link.aps.org/doi/10.1103/PhysRevLett.106.197001}.

\bibitem[{\citenamefont{Fang et~al.}(2011{\natexlab{a}})\citenamefont{Fang, Wu,
  Thomale, Bernevig, and Hu}}]{Fang2011_sWave}
\bibinfo{author}{\bibfnamefont{C.}~\bibnamefont{Fang}},
  \bibinfo{author}{\bibfnamefont{Y.-L.} \bibnamefont{Wu}},
  \bibinfo{author}{\bibfnamefont{R.}~\bibnamefont{Thomale}},
  \bibinfo{author}{\bibfnamefont{B.~A.} \bibnamefont{Bernevig}},
  \bibnamefont{and} \bibinfo{author}{\bibfnamefont{J.}~\bibnamefont{Hu}},
  \bibinfo{journal}{Phys. Rev. X} \textbf{\bibinfo{volume}{1}},
  \bibinfo{pages}{011009} (\bibinfo{year}{2011}{\natexlab{a}}),
  \urlprefix\url{http://link.aps.org/doi/10.1103/PhysRevX.1.011009}.

\bibitem[{\citenamefont{Wang et~al.}(2011{\natexlab{a}})\citenamefont{Wang,
  Song, Shi, Wang, Chen, Tian, Chen, Guo, Yang, and Li}}]{Z.Wang2011}
\bibinfo{author}{\bibfnamefont{Z.}~\bibnamefont{Wang}},
  \bibinfo{author}{\bibfnamefont{Y.~J.} \bibnamefont{Song}},
  \bibinfo{author}{\bibfnamefont{H.~L.} \bibnamefont{Shi}},
  \bibinfo{author}{\bibfnamefont{Z.~W.} \bibnamefont{Wang}},
  \bibinfo{author}{\bibfnamefont{Z.}~\bibnamefont{Chen}},
  \bibinfo{author}{\bibfnamefont{H.~F.} \bibnamefont{Tian}},
  \bibinfo{author}{\bibfnamefont{G.~F.} \bibnamefont{Chen}},
  \bibinfo{author}{\bibfnamefont{J.~G.} \bibnamefont{Guo}},
  \bibinfo{author}{\bibfnamefont{H.~X.} \bibnamefont{Yang}}, \bibnamefont{and}
  \bibinfo{author}{\bibfnamefont{J.~Q.} \bibnamefont{Li}},
  \bibinfo{journal}{Phys. Rev. B} \textbf{\bibinfo{volume}{83}},
  \bibinfo{pages}{140505} (\bibinfo{year}{2011}{\natexlab{a}}),
  \urlprefix\url{http://link.aps.org/doi/10.1103/PhysRevB.83.140505}.

\bibitem[{\citenamefont{Song et~al.}(2011)\citenamefont{Song, Wang, Wang, Shi,
  Chen, Tian, Chen, Yang, and Li}}]{Song2011}
\bibinfo{author}{\bibfnamefont{Y.~J.} \bibnamefont{Song}},
  \bibinfo{author}{\bibfnamefont{Z.}~\bibnamefont{Wang}},
  \bibinfo{author}{\bibfnamefont{Z.~W.} \bibnamefont{Wang}},
  \bibinfo{author}{\bibfnamefont{H.~L.} \bibnamefont{Shi}},
  \bibinfo{author}{\bibfnamefont{Z.}~\bibnamefont{Chen}},
  \bibinfo{author}{\bibfnamefont{H.~F.} \bibnamefont{Tian}},
  \bibinfo{author}{\bibfnamefont{G.~F.} \bibnamefont{Chen}},
  \bibinfo{author}{\bibfnamefont{H.~X.} \bibnamefont{Yang}}, \bibnamefont{and}
  \bibinfo{author}{\bibfnamefont{J.~Q.} \bibnamefont{Li}}, \bibinfo{journal}{E
  P L} \textbf{\bibinfo{volume}{95}}, \bibinfo{pages}{37007}
  (\bibinfo{year}{2011}).

\bibitem[{\citenamefont{Ricci et~al.}(2011)\citenamefont{Ricci, Poccia, Campi,
  Joseph, Arrighetti, Barba, Reynolds, Burghammer, Takeya, Mizuguchi
  et~al.}}]{Ricci2011}
\bibinfo{author}{\bibfnamefont{A.}~\bibnamefont{Ricci}},
  \bibinfo{author}{\bibfnamefont{N.}~\bibnamefont{Poccia}},
  \bibinfo{author}{\bibfnamefont{G.}~\bibnamefont{Campi}},
  \bibinfo{author}{\bibfnamefont{B.}~\bibnamefont{Joseph}},
  \bibinfo{author}{\bibfnamefont{G.}~\bibnamefont{Arrighetti}},
  \bibinfo{author}{\bibfnamefont{L.}~\bibnamefont{Barba}},
  \bibinfo{author}{\bibfnamefont{M.}~\bibnamefont{Reynolds}},
  \bibinfo{author}{\bibfnamefont{M.}~\bibnamefont{Burghammer}},
  \bibinfo{author}{\bibfnamefont{H.}~\bibnamefont{Takeya}},
  \bibinfo{author}{\bibfnamefont{Y.}~\bibnamefont{Mizuguchi}},
  \bibnamefont{et~al.}, \bibinfo{journal}{Phys. Rev. B}
  \textbf{\bibinfo{volume}{84}}, \bibinfo{pages}{060511}
  (\bibinfo{year}{2011}),
  \urlprefix\url{http://link.aps.org/doi/10.1103/PhysRevB.84.060511}.

\bibitem[{\citenamefont{Ksenofontov et~al.}(2011)\citenamefont{Ksenofontov,
  Wortmann, Medvedev, Tsurkan, Deisenhofer, Loidl, and Felser}}]{Felser2011}
\bibinfo{author}{\bibfnamefont{V.}~\bibnamefont{Ksenofontov}},
  \bibinfo{author}{\bibfnamefont{G.}~\bibnamefont{Wortmann}},
  \bibinfo{author}{\bibfnamefont{S.~A.} \bibnamefont{Medvedev}},
  \bibinfo{author}{\bibfnamefont{V.}~\bibnamefont{Tsurkan}},
  \bibinfo{author}{\bibfnamefont{J.}~\bibnamefont{Deisenhofer}},
  \bibinfo{author}{\bibfnamefont{A.}~\bibnamefont{Loidl}}, \bibnamefont{and}
  \bibinfo{author}{\bibfnamefont{C.}~\bibnamefont{Felser}},
  \bibinfo{journal}{Phys. Rev. B} \textbf{\bibinfo{volume}{84}},
  \bibinfo{pages}{180508} (\bibinfo{year}{2011}),
  \urlprefix\url{http://link.aps.org/doi/10.1103/PhysRevB.84.180508}.

\bibitem[{\citenamefont{Yuan et~al.}(2012)\citenamefont{Yuan, Dong, Song,
  Zheng, Chen, Hu, Li, and Wang}}]{Yuan2012}
\bibinfo{author}{\bibfnamefont{R.~H.} \bibnamefont{Yuan}},
  \bibinfo{author}{\bibfnamefont{T.}~\bibnamefont{Dong}},
  \bibinfo{author}{\bibfnamefont{Y.~J.} \bibnamefont{Song}},
  \bibinfo{author}{\bibfnamefont{P.}~\bibnamefont{Zheng}},
  \bibinfo{author}{\bibfnamefont{G.~F.} \bibnamefont{Chen}},
  \bibinfo{author}{\bibfnamefont{J.~P.} \bibnamefont{Hu}},
  \bibinfo{author}{\bibfnamefont{J.~Q.} \bibnamefont{Li}}, \bibnamefont{and}
  \bibinfo{author}{\bibfnamefont{N.~L.} \bibnamefont{Wang}},
  \bibinfo{journal}{Sci. Rep.} \textbf{\bibinfo{volume}{2}},
  \bibinfo{pages}{221} (\bibinfo{year}{2012}).

\bibitem[{\citenamefont{Li et~al.}(2012{\natexlab{a}})\citenamefont{Li, Ding,
  Deng, Chang, Song, He, Wang, Ma, Hu, Chen et~al.}}]{Li2011}
\bibinfo{author}{\bibfnamefont{W.}~\bibnamefont{Li}},
  \bibinfo{author}{\bibfnamefont{H.}~\bibnamefont{Ding}},
  \bibinfo{author}{\bibfnamefont{P.}~\bibnamefont{Deng}},
  \bibinfo{author}{\bibfnamefont{K.}~\bibnamefont{Chang}},
  \bibinfo{author}{\bibfnamefont{C.}~\bibnamefont{Song}},
  \bibinfo{author}{\bibfnamefont{K.}~\bibnamefont{He}},
  \bibinfo{author}{\bibfnamefont{L.}~\bibnamefont{Wang}},
  \bibinfo{author}{\bibfnamefont{X.}~\bibnamefont{Ma}},
  \bibinfo{author}{\bibfnamefont{J.-P.} \bibnamefont{Hu}},
  \bibinfo{author}{\bibfnamefont{X.}~\bibnamefont{Chen}}, \bibnamefont{et~al.},
  \bibinfo{journal}{Nature Phys.} \textbf{\bibinfo{volume}{8}},
  \bibinfo{pages}{126} (\bibinfo{year}{2012}{\natexlab{a}}).

\bibitem[{\citenamefont{Ma et~al.}(2011)\citenamefont{Ma, Ji, Dai, He, Wang,
  Chen, Normand, and Yu}}]{Ma2011}
\bibinfo{author}{\bibfnamefont{L.}~\bibnamefont{Ma}},
  \bibinfo{author}{\bibfnamefont{G.~F.} \bibnamefont{Ji}},
  \bibinfo{author}{\bibfnamefont{J.}~\bibnamefont{Dai}},
  \bibinfo{author}{\bibfnamefont{J.~B.} \bibnamefont{He}},
  \bibinfo{author}{\bibfnamefont{D.~M.} \bibnamefont{Wang}},
  \bibinfo{author}{\bibfnamefont{G.~F.} \bibnamefont{Chen}},
  \bibinfo{author}{\bibfnamefont{B.}~\bibnamefont{Normand}}, \bibnamefont{and}
  \bibinfo{author}{\bibfnamefont{W.}~\bibnamefont{Yu}}, \bibinfo{journal}{Phys.
  Rev. B} \textbf{\bibinfo{volume}{84}}, \bibinfo{pages}{220505}
  (\bibinfo{year}{2011}),
  \urlprefix\url{http://link.aps.org/doi/10.1103/PhysRevB.84.220505}.

\bibitem[{\citenamefont{Charnukha et~al.}(2012)\citenamefont{Charnukha,
  Cvitkovic, Prokscha, Pr\"opper, Ocelic, Suter, Salman, Morenzoni,
  Deisenhofer, Tsurkan et~al.}}]{Charnukha2012}
\bibinfo{author}{\bibfnamefont{A.}~\bibnamefont{Charnukha}},
  \bibinfo{author}{\bibfnamefont{A.}~\bibnamefont{Cvitkovic}},
  \bibinfo{author}{\bibfnamefont{T.}~\bibnamefont{Prokscha}},
  \bibinfo{author}{\bibfnamefont{D.}~\bibnamefont{Pr\"opper}},
  \bibinfo{author}{\bibfnamefont{N.}~\bibnamefont{Ocelic}},
  \bibinfo{author}{\bibfnamefont{A.}~\bibnamefont{Suter}},
  \bibinfo{author}{\bibfnamefont{Z.}~\bibnamefont{Salman}},
  \bibinfo{author}{\bibfnamefont{E.}~\bibnamefont{Morenzoni}},
  \bibinfo{author}{\bibfnamefont{J.}~\bibnamefont{Deisenhofer}},
  \bibinfo{author}{\bibfnamefont{V.}~\bibnamefont{Tsurkan}},
  \bibnamefont{et~al.}, \bibinfo{journal}{Phys. Rev. Lett.}
  \textbf{\bibinfo{volume}{109}}, \bibinfo{pages}{017003}
  (\bibinfo{year}{2012}),
  \urlprefix\url{http://link.aps.org/doi/10.1103/PhysRevLett.109.017003}.

\bibitem[{\citenamefont{Guo et~al.}(2010)\citenamefont{Guo, Jin, Wang, Wang,
  Zhu, Zhou, He, and Chen}}]{Guo2010}
\bibinfo{author}{\bibfnamefont{J.}~\bibnamefont{Guo}},
  \bibinfo{author}{\bibfnamefont{S.}~\bibnamefont{Jin}},
  \bibinfo{author}{\bibfnamefont{G.}~\bibnamefont{Wang}},
  \bibinfo{author}{\bibfnamefont{S.}~\bibnamefont{Wang}},
  \bibinfo{author}{\bibfnamefont{K.}~\bibnamefont{Zhu}},
  \bibinfo{author}{\bibfnamefont{T.}~\bibnamefont{Zhou}},
  \bibinfo{author}{\bibfnamefont{M.}~\bibnamefont{He}}, \bibnamefont{and}
  \bibinfo{author}{\bibfnamefont{X.}~\bibnamefont{Chen}},
  \bibinfo{journal}{Phys. Rev. B} \textbf{\bibinfo{volume}{82}},
  \bibinfo{pages}{180520} (\bibinfo{year}{2010}),
  \urlprefix\url{http://link.aps.org/doi/10.1103/PhysRevB.82.180520}.

\bibitem[{\citenamefont{Wang et~al.}(2011{\natexlab{b}})\citenamefont{Wang,
  Ying, Yan, Liu, Luo, Li, Wang, Zhang, Ye, Cheng et~al.}}]{Wang2011}
\bibinfo{author}{\bibfnamefont{A.~F.} \bibnamefont{Wang}},
  \bibinfo{author}{\bibfnamefont{J.~J.} \bibnamefont{Ying}},
  \bibinfo{author}{\bibfnamefont{Y.~J.} \bibnamefont{Yan}},
  \bibinfo{author}{\bibfnamefont{R.~H.} \bibnamefont{Liu}},
  \bibinfo{author}{\bibfnamefont{X.~G.} \bibnamefont{Luo}},
  \bibinfo{author}{\bibfnamefont{Z.~Y.} \bibnamefont{Li}},
  \bibinfo{author}{\bibfnamefont{X.~F.} \bibnamefont{Wang}},
  \bibinfo{author}{\bibfnamefont{M.}~\bibnamefont{Zhang}},
  \bibinfo{author}{\bibfnamefont{G.~J.} \bibnamefont{Ye}},
  \bibinfo{author}{\bibfnamefont{P.}~\bibnamefont{Cheng}},
  \bibnamefont{et~al.}, \bibinfo{journal}{Phys. Rev. B}
  \textbf{\bibinfo{volume}{83}}, \bibinfo{pages}{060512}
  (\bibinfo{year}{2011}{\natexlab{b}}),
  \urlprefix\url{http://link.aps.org/doi/10.1103/PhysRevB.83.060512}.

\bibitem[{\citenamefont{Krzton-Maziopa
  et~al.}(2011)\citenamefont{Krzton-Maziopa, Shermadini, Pomjakushina,
  Pomjakushin, Bendele, Amato, Khasanov, Luetkens, and Conder}}]{Krzton2011}
\bibinfo{author}{\bibfnamefont{A.}~\bibnamefont{Krzton-Maziopa}},
  \bibinfo{author}{\bibfnamefont{Z.}~\bibnamefont{Shermadini}},
  \bibinfo{author}{\bibfnamefont{E.}~\bibnamefont{Pomjakushina}},
  \bibinfo{author}{\bibfnamefont{V.}~\bibnamefont{Pomjakushin}},
  \bibinfo{author}{\bibfnamefont{M.}~\bibnamefont{Bendele}},
  \bibinfo{author}{\bibfnamefont{A.}~\bibnamefont{Amato}},
  \bibinfo{author}{\bibfnamefont{R.}~\bibnamefont{Khasanov}},
  \bibinfo{author}{\bibfnamefont{H.}~\bibnamefont{Luetkens}}, \bibnamefont{and}
  \bibinfo{author}{\bibfnamefont{K.}~\bibnamefont{Conder}},
  \bibinfo{journal}{J. Phys. Condens. Matter} \textbf{\bibinfo{volume}{23}},
  \bibinfo{pages}{052203} (\bibinfo{year}{2011}).

\bibitem[{\citenamefont{Fang et~al.}(2011{\natexlab{b}})\citenamefont{Fang,
  Wang, Dong, Li, Feng, Chen, and Yuan}}]{Fang2011}
\bibinfo{author}{\bibfnamefont{M.~H.} \bibnamefont{Fang}},
  \bibinfo{author}{\bibfnamefont{H.~D.} \bibnamefont{Wang}},
  \bibinfo{author}{\bibfnamefont{C.~H.} \bibnamefont{Dong}},
  \bibinfo{author}{\bibfnamefont{Z.~J.} \bibnamefont{Li}},
  \bibinfo{author}{\bibfnamefont{C.~M.} \bibnamefont{Feng}},
  \bibinfo{author}{\bibfnamefont{J.}~\bibnamefont{Chen}}, \bibnamefont{and}
  \bibinfo{author}{\bibfnamefont{H.~Q.} \bibnamefont{Yuan}},
  \bibinfo{journal}{Europhys. Lett.} \textbf{\bibinfo{volume}{94}},
  \bibinfo{pages}{27009} (\bibinfo{year}{2011}{\natexlab{b}}).

\bibitem[{\citenamefont{Bao et~al.}()\citenamefont{Bao, Li, Huang, Chen, He,
  Green, Qiu, Wang, and Luo}}]{Bao2011b}
\bibinfo{author}{\bibfnamefont{W.}~\bibnamefont{Bao}},
  \bibinfo{author}{\bibfnamefont{G.~N.} \bibnamefont{Li}},
  \bibinfo{author}{\bibfnamefont{Q.}~\bibnamefont{Huang}},
  \bibinfo{author}{\bibfnamefont{G.~F.} \bibnamefont{Chen}},
  \bibinfo{author}{\bibfnamefont{J.~B.} \bibnamefont{He}},
  \bibinfo{author}{\bibfnamefont{M.~A.} \bibnamefont{Green}},
  \bibinfo{author}{\bibfnamefont{Y.}~\bibnamefont{Qiu}},
  \bibinfo{author}{\bibfnamefont{D.~M.} \bibnamefont{Wang}}, \bibnamefont{and}
  \bibinfo{author}{\bibfnamefont{J.~L.} \bibnamefont{Luo}},
  \bibinfo{journal}{arXiv:1102.3674v1}  (????).

\bibitem[{\citenamefont{Wang et~al.}(2011{\natexlab{c}})\citenamefont{Wang, He,
  Xia, and Chen}}]{D.Wang2011}
\bibinfo{author}{\bibfnamefont{D.~M.} \bibnamefont{Wang}},
  \bibinfo{author}{\bibfnamefont{J.~B.} \bibnamefont{He}},
  \bibinfo{author}{\bibfnamefont{T.-L.} \bibnamefont{Xia}}, \bibnamefont{and}
  \bibinfo{author}{\bibfnamefont{G.~F.} \bibnamefont{Chen}},
  \bibinfo{journal}{Phys. Rev. B} \textbf{\bibinfo{volume}{83}},
  \bibinfo{pages}{132502} (\bibinfo{year}{2011}{\natexlab{c}}),
  \urlprefix\url{http://link.aps.org/doi/10.1103/PhysRevB.83.132502}.

\bibitem[{\citenamefont{Texier et~al.}(2012)\citenamefont{Texier, Deisenhofer,
  Tsurkan, Loidl, Inosov, Friemel, and Bobroff}}]{Texier2012}
\bibinfo{author}{\bibfnamefont{Y.}~\bibnamefont{Texier}},
  \bibinfo{author}{\bibfnamefont{J.}~\bibnamefont{Deisenhofer}},
  \bibinfo{author}{\bibfnamefont{V.}~\bibnamefont{Tsurkan}},
  \bibinfo{author}{\bibfnamefont{A.}~\bibnamefont{Loidl}},
  \bibinfo{author}{\bibfnamefont{D.~S.} \bibnamefont{Inosov}},
  \bibinfo{author}{\bibfnamefont{G.}~\bibnamefont{Friemel}}, \bibnamefont{and}
  \bibinfo{author}{\bibfnamefont{J.}~\bibnamefont{Bobroff}},
  \bibinfo{journal}{Phys. Rev. Lett.} \textbf{\bibinfo{volume}{108}},
  \bibinfo{pages}{237002} (\bibinfo{year}{2012}),
  \urlprefix\url{http://link.aps.org/doi/10.1103/PhysRevLett.108.237002}.

\bibitem[{\citenamefont{Li et~al.}(2012{\natexlab{b}})\citenamefont{Li, Ding,
  Li, Deng, Chang, He, Ji, Wang, Ma, Hu et~al.}}]{Li2012}
\bibinfo{author}{\bibfnamefont{W.}~\bibnamefont{Li}},
  \bibinfo{author}{\bibfnamefont{H.}~\bibnamefont{Ding}},
  \bibinfo{author}{\bibfnamefont{Z.}~\bibnamefont{Li}},
  \bibinfo{author}{\bibfnamefont{P.}~\bibnamefont{Deng}},
  \bibinfo{author}{\bibfnamefont{K.}~\bibnamefont{Chang}},
  \bibinfo{author}{\bibfnamefont{K.}~\bibnamefont{He}},
  \bibinfo{author}{\bibfnamefont{S.}~\bibnamefont{Ji}},
  \bibinfo{author}{\bibfnamefont{L.}~\bibnamefont{Wang}},
  \bibinfo{author}{\bibfnamefont{X.}~\bibnamefont{Ma}},
  \bibinfo{author}{\bibfnamefont{J.-P.} \bibnamefont{Hu}},
  \bibnamefont{et~al.}, \bibinfo{journal}{Phys. Rev. Lett.}
  \textbf{\bibinfo{volume}{109}}, \bibinfo{pages}{057003}
  (\bibinfo{year}{2012}{\natexlab{b}}),
  \urlprefix\url{http://link.aps.org/doi/10.1103/PhysRevLett.109.057003}.

\bibitem[{\citenamefont{Friemel et~al.}(2012)\citenamefont{Friemel, Park,
  Maier, Tsurkan, Li, Deisenhofer, Krug~von Nidda, Loidl, Ivanov, Keimer
  et~al.}}]{PhysRevB.85.140511}
\bibinfo{author}{\bibfnamefont{G.}~\bibnamefont{Friemel}},
  \bibinfo{author}{\bibfnamefont{J.~T.} \bibnamefont{Park}},
  \bibinfo{author}{\bibfnamefont{T.~A.} \bibnamefont{Maier}},
  \bibinfo{author}{\bibfnamefont{V.}~\bibnamefont{Tsurkan}},
  \bibinfo{author}{\bibfnamefont{Y.}~\bibnamefont{Li}},
  \bibinfo{author}{\bibfnamefont{J.}~\bibnamefont{Deisenhofer}},
  \bibinfo{author}{\bibfnamefont{H.-A.} \bibnamefont{Krug~von Nidda}},
  \bibinfo{author}{\bibfnamefont{A.}~\bibnamefont{Loidl}},
  \bibinfo{author}{\bibfnamefont{A.}~\bibnamefont{Ivanov}},
  \bibinfo{author}{\bibfnamefont{B.}~\bibnamefont{Keimer}},
  \bibnamefont{et~al.}, \bibinfo{journal}{Phys. Rev. B}
  \textbf{\bibinfo{volume}{85}}, \bibinfo{pages}{140511}
  (\bibinfo{year}{2012}),
  \urlprefix\url{http://link.aps.org/doi/10.1103/PhysRevB.85.140511}.

\bibitem[{\citenamefont{Han et~al.}(2012)\citenamefont{Han, Yang, Shen, Wang,
  Li, and Wen}}]{Han2011}
\bibinfo{author}{\bibfnamefont{F.}~\bibnamefont{Han}},
  \bibinfo{author}{\bibfnamefont{H.}~\bibnamefont{Yang}},
  \bibinfo{author}{\bibfnamefont{B.}~\bibnamefont{Shen}},
  \bibinfo{author}{\bibfnamefont{Z.-Y.} \bibnamefont{Wang}},
  \bibinfo{author}{\bibfnamefont{C.~H.} \bibnamefont{Li}}, \bibnamefont{and}
  \bibinfo{author}{\bibfnamefont{H.-H.} \bibnamefont{Wen}},
  \bibinfo{journal}{Philos. Mag.} \textbf{\bibinfo{volume}{92}},
  \bibinfo{pages}{2553} (\bibinfo{year}{2012}).

\bibitem[{\citenamefont{Zavalij et~al.}(2011)\citenamefont{Zavalij, Bao, Wang,
  Ying, Chen, Wang, He, Wang, Chen, Hsieh et~al.}}]{Zavalij2011}
\bibinfo{author}{\bibfnamefont{P.}~\bibnamefont{Zavalij}},
  \bibinfo{author}{\bibfnamefont{W.}~\bibnamefont{Bao}},
  \bibinfo{author}{\bibfnamefont{X.~F.} \bibnamefont{Wang}},
  \bibinfo{author}{\bibfnamefont{J.~J.} \bibnamefont{Ying}},
  \bibinfo{author}{\bibfnamefont{X.~H.} \bibnamefont{Chen}},
  \bibinfo{author}{\bibfnamefont{D.~M.} \bibnamefont{Wang}},
  \bibinfo{author}{\bibfnamefont{J.~B.} \bibnamefont{He}},
  \bibinfo{author}{\bibfnamefont{X.~Q.} \bibnamefont{Wang}},
  \bibinfo{author}{\bibfnamefont{G.~F.} \bibnamefont{Chen}},
  \bibinfo{author}{\bibfnamefont{P.-Y.} \bibnamefont{Hsieh}},
  \bibnamefont{et~al.}, \bibinfo{journal}{Phys. Rev. B}
  \textbf{\bibinfo{volume}{83}}, \bibinfo{pages}{132509}
  (\bibinfo{year}{2011}),
  \urlprefix\url{http://link.aps.org/doi/10.1103/PhysRevB.83.132509}.

\bibitem[{\citenamefont{Pomjakushin et~al.}(2011)\citenamefont{Pomjakushin,
  Sheptyakov, Pomjakushina, Krzton-Maziopa, Conder, Chernyshov, Svitlyk, and
  Shermadini}}]{Pomjakushin2011}
\bibinfo{author}{\bibfnamefont{V.~Y.} \bibnamefont{Pomjakushin}},
  \bibinfo{author}{\bibfnamefont{D.~V.} \bibnamefont{Sheptyakov}},
  \bibinfo{author}{\bibfnamefont{E.~V.} \bibnamefont{Pomjakushina}},
  \bibinfo{author}{\bibfnamefont{A.}~\bibnamefont{Krzton-Maziopa}},
  \bibinfo{author}{\bibfnamefont{K.}~\bibnamefont{Conder}},
  \bibinfo{author}{\bibfnamefont{D.}~\bibnamefont{Chernyshov}},
  \bibinfo{author}{\bibfnamefont{V.}~\bibnamefont{Svitlyk}}, \bibnamefont{and}
  \bibinfo{author}{\bibfnamefont{Z.}~\bibnamefont{Shermadini}},
  \bibinfo{journal}{Phys. Rev. B} \textbf{\bibinfo{volume}{83}},
  \bibinfo{pages}{144410} (\bibinfo{year}{2011}),
  \urlprefix\url{http://link.aps.org/doi/10.1103/PhysRevB.83.144410}.

\bibitem[{\citenamefont{Sales et~al.}(2011)\citenamefont{Sales, McGuire, May,
  Cao, Chakoumakos, and Sefat}}]{SalesTFS}
\bibinfo{author}{\bibfnamefont{B.~C.} \bibnamefont{Sales}},
  \bibinfo{author}{\bibfnamefont{M.~A.} \bibnamefont{McGuire}},
  \bibinfo{author}{\bibfnamefont{A.~F.} \bibnamefont{May}},
  \bibinfo{author}{\bibfnamefont{H.}~\bibnamefont{Cao}},
  \bibinfo{author}{\bibfnamefont{B.~C.} \bibnamefont{Chakoumakos}},
  \bibnamefont{and} \bibinfo{author}{\bibfnamefont{A.~S.} \bibnamefont{Sefat}},
  \bibinfo{journal}{Phys. Rev. B} \textbf{\bibinfo{volume}{83}},
  \bibinfo{pages}{224510} (\bibinfo{year}{2011}).

\bibitem[{\citenamefont{Cao et~al.}(2012)\citenamefont{Cao, Cantoni, May,
  McGuire, Chakoumakos, Pennycook, Custelcean, Sefat, and Sales}}]{CaoTFS}
\bibinfo{author}{\bibfnamefont{H.}~\bibnamefont{Cao}},
  \bibinfo{author}{\bibfnamefont{C.}~\bibnamefont{Cantoni}},
  \bibinfo{author}{\bibfnamefont{A.~F.} \bibnamefont{May}},
  \bibinfo{author}{\bibfnamefont{M.~A.} \bibnamefont{McGuire}},
  \bibinfo{author}{\bibfnamefont{B.~C.} \bibnamefont{Chakoumakos}},
  \bibinfo{author}{\bibfnamefont{S.~J.} \bibnamefont{Pennycook}},
  \bibinfo{author}{\bibfnamefont{R.}~\bibnamefont{Custelcean}},
  \bibinfo{author}{\bibfnamefont{A.~S.} \bibnamefont{Sefat}}, \bibnamefont{and}
  \bibinfo{author}{\bibfnamefont{B.~C.} \bibnamefont{Sales}},
  \bibinfo{journal}{Phys. Rev. B} \textbf{\bibinfo{volume}{85}},
  \bibinfo{pages}{054515} (\bibinfo{year}{2012}),
  \urlprefix\url{http://link.aps.org/doi/10.1103/PhysRevB.85.054515}.

\bibitem[{\citenamefont{R\"ohlsberger}(2010)}]{Rohlsberger}
\bibinfo{author}{\bibfnamefont{R.}~\bibnamefont{R\"ohlsberger}},
  \emph{\bibinfo{title}{Nuclear Condensed Matter Physics with Synchrotron
  Radiation: Basic Principles, Methodology and Applications}}
  (\bibinfo{publisher}{Springer}, \bibinfo{address}{new York},
  \bibinfo{year}{2010}).

\bibitem[{\citenamefont{R$\ddot{\mathrm{u}}$ffer and Chumakov}(1996)}]{Ruffer}
\bibinfo{author}{\bibfnamefont{R.}~\bibnamefont{R$\ddot{\mathrm{u}}$ffer}}
  \bibnamefont{and} \bibinfo{author}{\bibfnamefont{A.~I.}
  \bibnamefont{Chumakov}}, \bibinfo{journal}{Hyperfine Interact.}
  \textbf{\bibinfo{volume}{97-98}}, \bibinfo{pages}{589}
  (\bibinfo{year}{1996}).

\bibitem[{\citenamefont{Shvyd'ko}(2000)}]{MOTIF}
\bibinfo{author}{\bibfnamefont{Y.~V.} \bibnamefont{Shvyd'ko}},
  \bibinfo{journal}{Hyperfine Interact.} \textbf{\bibinfo{volume}{125}},
  \bibinfo{pages}{173} (\bibinfo{year}{2000}).

\bibitem[{\citenamefont{Lumsden and
  Christianson}(2010)}]{Lumsden_Christianson_Review}
\bibinfo{author}{\bibfnamefont{M.~D.} \bibnamefont{Lumsden}} \bibnamefont{and}
  \bibinfo{author}{\bibfnamefont{A.~D.} \bibnamefont{Christianson}},
  \bibinfo{journal}{J. Phys. Condens. Matter} \textbf{\bibinfo{volume}{22}},
  \bibinfo{pages}{203203} (\bibinfo{year}{2010}).

\bibitem[{\citenamefont{Zhang and Singh}(2009)}]{ZhangSingh2009}
\bibinfo{author}{\bibfnamefont{L.}~\bibnamefont{Zhang}} \bibnamefont{and}
  \bibinfo{author}{\bibfnamefont{D.~J.} \bibnamefont{Singh}},
  \bibinfo{journal}{Phys. Rev. B} \textbf{\bibinfo{volume}{79}},
  \bibinfo{pages}{094528} (\bibinfo{year}{2009}).

\end{thebibliography}
\end{document}